\documentclass{elsart}

\begin{document}

\begin{frontmatter}

\title{On the structure of genealogical trees in the presence of
                 selection}

\author{P. R. A. Campos},
\author{M. T. Sonoda},
\author{J. F. Fontanari \thanksref{email}} 
\address{Instituto de F\'{\i}sica de S\~ao Carlos,
Universidade de S\~ao Paulo,
Caixa Postal 369,
13560-970 S\~ao Carlos, SP, Brazil}

\thanks[email]{Corresponding author. E-mail:
fontanari@if.sc.usp.br}

\begin{abstract}

We investigate through numerical simulations the effect of
selection on two summary statistics for nucleotide variation
in a sample of two genes from a 
population of $N$ asexually reproducing haploid individuals. One is
the mean time since two individuals had their most recent  common ancestor 
($\overline{T_s}$),  
and the other is the mean number of nucleotide differences between two 
genes
in the sample ($\overline{d_s}$).  In the case of diminishing
epistasis, in which the deleterious effect of a new mutation is attenuated,
we find that the scale of $\overline{d_s}$ with the population
size  depends on the  mutation rate,  leading then to the onset of
a sharp threshold phenomenon as $N$ becomes large.

\end{abstract}


\end{frontmatter}

\section{Introduction}\label{Intro}

The rapid accumulation
of DNA sequence data in the last two decades resulted in a 
shift of emphasis in population genetics 
from a prospective approach, which focuses on the changes in the
population composition with time, to a retrospective approach, which 
explores the
patterns of similarities between the different sequences to obtain
information about the evolutionary history of those sequences. Although
neutral genealogical processes have been the subject
of much attention in those decades, culminating with 
Kingman's Coalescent Theory \cite{Kingman}
(see \cite{Hudson,Fu} for reviews), very little is known about genealogical
processes with selection (see \cite{Kaplan,Higgs,Neuhauser} for a few exceptions).
The purpose of this paper is to study numerically the effect of selection
on two widely used summary statistics for nucleotide variation
in a random  sample of two genes, namely, the mean time 
since their most recent common ancestor, $\overline{T}_s$, 
and the average distance (measured
by the number of different nucleotides in homologous sites) between them,
$\overline{d}_s$. This last quantity is easily measured from DNA sequence data
and, within the neutral evolution assumption, can be used to estimate the product 
between the effective population size $N_e$ and the mutation rate per gene per generation
$U$ \cite{Hudson}. In this paper we show that this estimation procedure  
holds also in the case of diminishing epistasis, provided that  $U$ is not too small.

\section{Model}\label{Model}

We consider a haploid population of $N$ individuals or genes
that evolves 
according to the discrete-time Wright-Fisher model with selection 
and mutation \cite{Crow}. Here, haploid
means that each individual has only one copy of each chromosome as,
for instance, the mitochondrial genes which are inherited maternally.
In this sense, we will use interchangeably the term gene and individual
to refer to the unit of selection.
An individual is represented by an
infinite sequence of sites, each labeled $0$ or $1$: the bit $0$ denotes
the original (ancestral) type, and the bit $1$ a mutant type. 
The fitness of 
an individual
with $k$ mutations is $ w_k = \left ( 1 - s \right )^{k^\alpha}$ where 
$s \in \left ( 0, 1 \right )$ is the selective advantage per site of  
the original  nucleotide type, and 
$\alpha \geq 0$ is the epistasis parameter. The case $\alpha = 1$
corresponds to absence of epistasis, i.e., each new mutation reduces
the fitness of the individual by the same amount, irrespective of the 
number of previous mutations. The case $\alpha > 1$ 
(synergistic epistasis) models the situation
where the disadvantageous effect of a new mutation increases with the
number of mutations already present, while the case $\alpha < 1$
(diminishing epistasis) corresponds to the situation where the 
deleterious effect of a new mutation is attenuated.
The mutation mechanism is such that a mutant offspring gets a
mutation at a single new site that has never before seen a mutation.
In particular, we assume
that the probability that $k$ new mutations occur in one individual is given
by the Poisson distribution
\begin{equation}\label{M} 
M_k = \mbox{e}^{- U } ~\frac{U^k}{k!} ,
\end{equation}
where $U$ is the
mean number of new mutations per individual per generation.
The relevant quantities that characterize the population are
measured after the procedures of selection and mutation, 
in that order. The model presented above is the celebrated
infinite-sites model \cite{Kim71,Watterson} which 
has been widely used by population geneticists to describe
the DNA variability observed in samples of genes in the case of
neutral mutations ($s=0$).

\section{Analytical results}\label{Analytical}

In the  neutral limit ($s = 0$) as well as in the strong-selection 
limit ($s=1$) we can easily calculate  $\overline{T}_s$ and 
$\overline{d}_s$ analytically \cite{Higgs,DP}. Clearly, the value of the 
epistasis parameter is irrelevant in those limits. 

We  consider first the neutral limit. 
Let $T_{\alpha \beta}$ be the time in generations since the latest common ancestor
of individuals $\alpha$ and $\beta$. In the following we will calculate the probability
$\overline{P}_0(T)$ that two randomly chosen individuals have $T_{\alpha \beta} = T$.
The notation $\overline{ \left ( \ldots \right ) }$ stands for an average over
independent populations.
The probability that two individuals have no common ancestor in the preceding
generation is simply $\left ( 1 - 1/N \right ) $. So for $N$ large
the probability that their ancestors, $t$
generations ago, are all different is $ \mbox{e}^{-t/N}$.
Hence  the probability that the latest common ancestor of the two 
individuals lived exactly at 
$T$ generations ago is given by
\begin{equation}\label{anc}
\overline{P}_0 \left ( T \right ) = \left [ 1 -  \mbox{e}^{-(T+1)/N} \right ] -  
\left [ 1 -  \mbox{e}^{-T/N} \right ] \approx \frac{\mbox{e}^{-T/N}}{N} ,
\end{equation}
from where we obtain $\overline{T}_0 = N$. 
The relevant time scale in the neutral case is thus proportional to $N$.
The probability distribution
$\overline{P}_s \left ( T \right )$, which determines the statistical properties
of the genealogies, depends on such factors as the population size,
geographic structure and the distribution of fitness among the individuals.
It should be stressed that neutral mutations (i.e., mutations that do not
affect the fitness of the individuals) have no effect on the genealogies
of random samples \cite{Hudson} and so Eqn.\ (\ref{anc}) holds true
irrespective of the value of the mutation rate $U$. Of course,
the distance (the number of different nucleotides)
between two sampled sequences
depends strongly on the mutation process. For instance, for $U = 0$ 
all sequences in the sample are identical.
In the sequel we will calculate analytically the distribution of
distances between two sampled sequences.

We assume that during each time interval $dt$ 
\footnote{This continuous-time formulation yields the same results as the discrete-time 
model presented before in the limit of large $N$,  since in that case the 
relevant time scale is of order of $N$ generations, and so it is much larger than 
the time unit, i.e.,  one generation.}
each sequence has a probability 
$U dt$ of mutating to a new one that has never before been present
in the population. Thus, in the neutral case the probability 
$\phi_n^0 (t)$ that a sequence
differs from  its ancestral on $n$ sites  
after the divergence time 
$t$ obeys the equation
\begin{equation}\label{dif}
\frac{d \phi_n^0 (t)}{dt} = U \left [ \phi_{n-1}^0 (t)
- \phi_n^0 (t) \right ]
\end{equation}
whose solution is the Poisson distribution
\begin{equation}\label{phif}
\phi_n^0 (t) = \mbox{e}^{-U t} \, \frac{ \left ( Ut \right )^n}{n!} .
\end{equation}
Hence the average distance  between an individual
and its ancestor increases linearly with time $\overline{n}_ 0 = U t$.
In a more general context the steady accumulation of unfavorable mutations
in an asexual population is referred to as the  Muller's rachet \cite{Higgs}.

Let $\overline{W}_n^0$ be the probability that the distance between two sampled
individuals is equal $n$. Since in an asexual population the individuals
are all descended from a common ancestor at some point in the past 
we can write
\begin{equation}
\overline{W}_n^0 = \int_0^\infty dT \overline{P}_0 (T) \phi_n^0 (2T) .
\end{equation}
Using Eqns.\ (\ref{anc}) and (\ref{phif}), the integral can easily be evaluated yielding 
\begin{equation}\label{d_fin}
\overline{W}_n^0 = \frac{\lambda}{ \left ( \lambda + 1 \right )^{n+1} } 
\end{equation}
where $ \lambda = 1/2UN$. Hence, $\overline{d}_0 = 1/\lambda = 2 U N $.

We turn now to the analysis of the strong-selection limit 
($s =1$). Since only individuals with $k=0$ mutations can
generate offspring there is no difference in the fitness of the breeding
individuals. This limit is similar to the neutral limit in the sense that
the probability of two individuals having the same parent is $1/N_0$ where
$N_0$ is the number of individuals with $k=0$ mutations. Clearly, at any generation
$N_0$ is random variable distributed according to the binomial distribution
\begin{equation}\label{binomial}
  B \left ( N_0 \right )  =   \left ( \! \! \begin{array}{c} N \\ N_0 \end{array}
      \! \! \right ) \,
      \mbox{e}^{-UN_0} \, \left  (1 - \mbox{e}^{-U}\right )^{N -N_0}.
\end{equation} 
If $U$ is of order $1$ and $N$ is large we have $N_0 \approx
\overline{N_0} =  N \mbox{e}^{-U}$ so that $\overline{T}_1 = N_0$.
The probability distribution of the distance $d_1$ between two sampled individuals 
is equally easy to obtain:  it  is simply given by $M_{2 d_1}$ so that 
$ \overline{d}_1 = 2 U $.
We must note that since the probability
of extinction in one generation is 
$  B \left ( 0 \right ) $, 
the population will ultimately become extinct in a time of order of 
$1/  B \left ( 0 \right ) $.
However, if $U$ is not of order of $N$ the extinction times become so 
large that these events cannot be observed in the simulations 
described in the sequel.

\section{Simulations}\label{Simulations}

At any given time we keep track of the number of mutations
$k_i$ on each individual $i = 1, \ldots, N$, as well as of the identity 
of their parents. This information allows us to obtain
the most recent common ancestor of any two individuals and also the distance
between them.
To create a new generation from a given one
we assume, as usual, that the number of offspring that each individual
contributes to the new generation is proportional to its relative fitness,
$w_i/\overline{w}$ where $\overline{w} = \sum_i w_i$ is the total fitness of 
the population.
The offspring has all the mutations of its parent plus a random number
$k$ of new mutations distributed according to the Poisson distribution
$M_k$ given by Eqn.\ (\ref{M}). The initial population is composed of
$N$ individuals without mutations whose evolution we follow through typically
$1000$ generations. We then go backward in time to determine the common ancestors 
of each pair of individuals. We typically average our results over $300$ independent
runs.

%
\begin{figure}
\vspace{9truecm}
\includegraphics{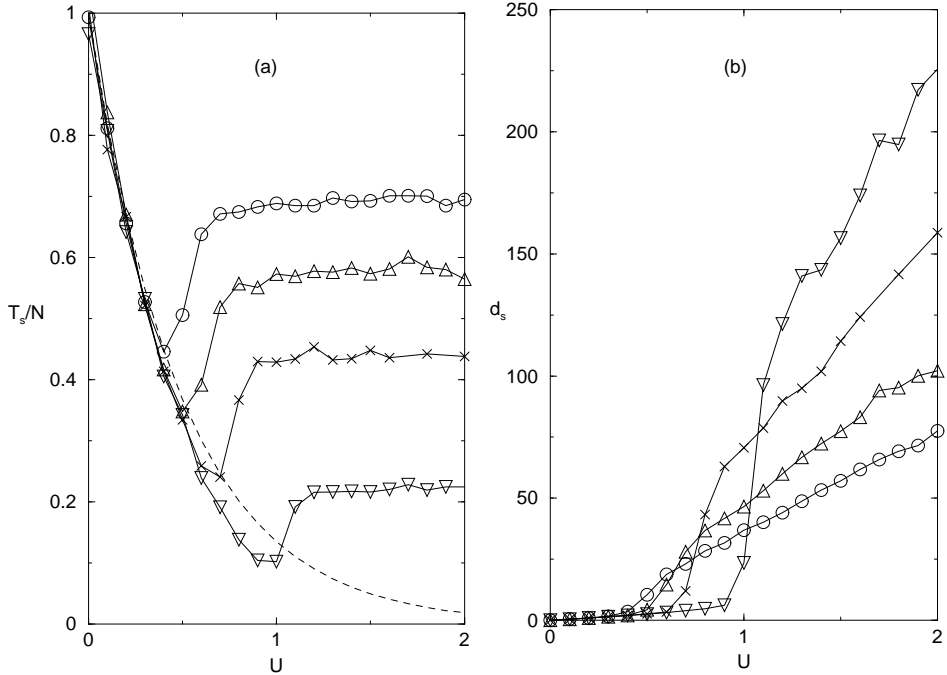}
\caption{$(a)$ Mean time since the most recent common ancestor of two
randomly chosen individuals as a function of the mutation rate $U$ for
$\alpha = s=0.5$. $(b)$ Average distance between two sampled individuals
as a function of $U$.
The convention is $N = 30~( \bigcirc )$, $ 50 ~(\bigtriangleup )$, $100 ~(\times )$,
and $300 ~(\bigtriangledown )$.
The dashed curve is $\exp (-U/s)$ while the solid lines are guides to the eyes.
}
\bigskip
\end{figure}

Here we present only the results for diminishing
epistasis ($\alpha = 0.5$). A more complete and detailed discussion 
will be presented elsewhere.
In Fig.\ 1  we show the dependence of
$\overline{T}_s$ and  $\overline{d}_s$ on
the mutation rate $U$ for $s=0.5$ and several values of the
population size $N$. 
One the one side, for small $U$ we find that both $\overline{T}_s/N$ and 
$\overline{d}_s$ are practically independent of $N$. In fact, 
similarly to the strong-selection limit,
in this regime we find $\overline{T}_s/N \approx \exp \left ( - U/s \right )$ and 
$\overline{d}_s \approx 2 U s $. 
On the other side,  for $U$ large we find a regime reminiscent
of the neutral limit, in which $\overline{T}_s/N$ becomes independent of 
the mutation rate. In this case we can define an effective population size
$N_e$, which depends on $s$ and $N$ but not on $U$, 
such that $\overline{T}_s = N_e$ and 
$\overline{d}_s \approx 2 U N_e $. 
As expected, we find that $N_e$ decreases with increasing $s$ since the number
of breeding individuals decreases with $s$. More specifically, 
$N_e$ seems to decrease like $N^{1-s}$ for $s$ not too close  $1$.
Interestingly, 
as illustrated in Fig. 1b the different scaling of $\overline{d}_s$ with $N$
in these two regimes leads to an abrupt increase of this quantity at a finite
value of $U$,  which may signal the existence of a phase transition
in the thermodynamic limit $N \rightarrow \infty$.
We note that these features are not observed for  $\alpha \geq 1$.



\noindent \textbf{Acknowledgments}

PRAC is supported by FAPESP and MTS is supported by CAPES.
The work of JFF was supported in part by CNPq. 
  


\end{document}